\documentclass[aps,preprint]{revtex4}%
\usepackage{amsfonts}
\usepackage{amsmath}
\usepackage{amssymb}
\usepackage{graphicx}%
\begin{document}
\title{{\LARGE BRIEF HISTORY OF BLACK-HOLES}}
\author{Marcelo Samuel Berman$^{(1)}$ }
\affiliation{$^{(1)}$Editora Albert Einstein Ltda.}
\affiliation{Av. Candido Hartman, 575 \# 17}
\affiliation{Ed. Renoir, b. Champagnat}
\affiliation{80730-440 Curitiba - PR - BRAZIL}
\affiliation{Fax \# (+5541) 342-9235}
\affiliation{eMail marsambe@yahoo.com}

\begin{abstract}
We show that the gravitational collapse of a black-hole terminates in the
birth of a white-hole, due to repulsive gravitation ( antigravitation\ ); in
particular, the infinite energy density singularity does NOT occur.

\end{abstract}
\maketitle

\bigskip

\bigskip

\bigskip

\bigskip

\bigskip

\bigskip

\bigskip

\bigskip

\begin{center}
{\large BRIEF HISTORY OF BLACK-HOLES}

\bigskip Marcelo Samuel Berman
\end{center}

\bigskip

\bigskip

We shall show that gravitational collapse endpoint is the birth of a white-hole.

\bigskip

When a certain charged rotating body enters in gravitational collapse, a
Kerr-Newman black-hole is formed$^{(1)}$ . Only three parameters define this
\textquotedblright hole\textquotedblright: its mass M, charge Q, and
rotational parameter \textquotedblright a\textquotedblright, such that
$a=-\frac{J}{M}$ where J is its total angular-momentum.

\bigskip

The energy contents, within a \textquotedblright radial\textquotedblright%
\ distance $\varrho$\ , of a Kerr-Newmann b.h., may be found from the
relation$^{(2)}$: \ \ (in G = c = 1 units)

\bigskip$E=M-\frac{\left(  Q^{2}+M^{2}\right)  }{4\varrho}\left[
1+\frac{\left(  a^{2}+\varrho^{2}\right)  }{a\varrho}arctgh\left(  \frac
{a}{\varrho}\right)  \right]  $ \ \ \ \ \ \ \ \ \ \ \ \ \ \ \ \ \ \ \ \ ,\ \ \ \ \ \ \ \ \ \ \ \ \ \ \ \ \ \ \ \ \ \ \ \ \ \ \ \ \ \ \ \ (1)

\bigskip

where $\varrho$ \ is the positive root of:

\bigskip

$\frac{x^{2}+y^{2}}{\varrho^{2}+a^{2}}+\frac{z^{2}}{\varrho^{2}}=1$ \ \ \ \ \ \ \ \ \ \ \ \ \ \ \ \ \ \ \ \ \ \ \ \ \ \ \ \ \ \ \ \ \ \ \ \ \ \ \ \ \ \ \ \ \ \ \ \ \ \ \ \ \ \ \ \ \ \ \ \ \ \ \ \ \ \ \ .\ \ \ \ \ \ \ \ \ \ \ \ \ \ \ \ \ \ \ \ \ \ \ \ (2)

\bigskip

While collapse is in process, $\varrho$ will become smaller, up to the point
when E becomes null, and, afterwards, negative. Berman$^{(2)}$ has suggested
that when E $\leq0$ ,  the  negative energy contents E within a certain
$\varrho$ value, is  characteristic of repulsive gravitation, or antigravity.

\bigskip

In such case, the collapse will be deaccelerated, and eventually halted and
reversed towards a radially increasing direction: the outcome is a white-hole.

\bigskip

The endpoint of gravitational collapse is, thus NOT a black-hole with singular
point of infinite energy density, but a surging \underline{white-hole}.The
positivity of total energy, which we define as the limit of E for infinite
radial distances,is \ assured,being given \ by M$c^{2},$ which is positive for
\ usual matter.

\bigskip

The time taken by the reversion of the collapse has to be dealt in a separate letter.

\bigskip

We can get a taste of the involved energy densities in the halting point by
obtaining the energy inside a spherical non-rotating \textquotedblright
ball\textquotedblright, (Q = a = 0 ), where,

\bigskip

$E=Mc^{2}-\frac{GM^{2}}{2R}$ \ \ \ \ \ \ \ \ \ \ \ \ \ \ \ \ \ \ \ \ \ \ \ \ \ \ \ \ \ \ \ \ \ \ \ \ \ \ \ \ \ \ \ \ \ \ \ \ \ \ \ \ \ \ \ \ \ \ \ \ \ \ \ \ \ \ \ \ \ \ \ \ \ \ \ \ \ \ \ \ \ \ \ \ \ \ \ \ \ \ \ \ \ \ \ \ (3)

\bigskip

We obtain,

\bigskip

$\mu=\frac{dE}{dV}=\frac{1}{4\pi R^{2}}\cdot\frac{dE}{dR}=\frac{GM^{2}}{8\pi
R^{4}}$ \ \ \ \ \ \ \ \ \ \ \ \ \ \ \ \ \ \ \ \ \ \ \ \ \ \ \ \ \ \ \ \ \ \ \ \ \ \ \ \ \ \ \ \ \ \ \ \ \ \ \ \ \ \ \ \ \ \ \ \ \ \ \ \ \ \ \ \ \ \ \ \ \ \ \ \ (4)

\bigskip

\bigskip

For E = E$_{0}=0$,

\bigskip

$R=R_{0}=\frac{GM}{2c^{2}}$ \ \ \ \ \ \ \ \ \ \ \ \ \ \ \ \ \ \ \ \ \ \ \ \ \ \ \ \ \ \ \ \ \ \ \ \ \ \ \ \ \ \ \ \ \ \ \ \ \ \ \ \ \ \ \ \ \ \ \ \ \ \ \ \ \ \ \ \ \ \ \ \ \ \ \ \ \ \ \ \ \ \ \ \ \ \ \ \ \ \ \ \ \ \ \ \ \ \ (5)

\bigskip

while

\bigskip

$\mu=\mu_{0}=\frac{2}{\pi}G^{-3}M^{-2}c^{8}$

\bigskip

The mass density is obtained by dividing $\ \mu\ \ $by$\ \ c^{2},$so,for the
Sun, $\ \ \ \ \ \mu_{0}/c^{2}\approx10^{21}$ g/cm$^{3}.$

\bigskip

The general result is, numerically,when M is given in grams,

$\mu_{0}/c^{2}\thickapprox10^{87}M^{-2}$ g/cm$^{3}.$

\bigskip

For the whole Universe,we estimate the very early mass as Planck's mass,
$\thickapprox10^{-5}$grams, so that the maximumm \ big-crunch \ mass density
would be $\thickapprox10^{97}$g/cm$^{3}.$

\bigskip

\bigskip

{\large Acknowledgements}

\bigskip

I acknowledge several conversation with Murari Mohan Som (\ M.M. Som ), of a
very enlightening nature, about this subject.

\bigskip

\bigskip

{\large References}

\bigskip

1. Newman, E. T.; et al. - Jrn. Math. Phys. \underline{6}, 918 ( 1965 ).

2. Berman, M.\ S. ( 2005 ) - \textquotedblright Energy of Black-Holes and
Hawking's Universe\textquotedblright,in \textit{Black-Holes:Research and
Development,}ed. F. Columbus, Nova Sc. Publ., N.Y., to be published in 2005.

\bigskip

\ \ 

\bigskip

\end{document}